\documentclass[prl,twocolumn,showpacs,superscriptaddress]{revtex4-1}

\usepackage{amsmath}
\usepackage{amssymb}

\usepackage{longtable}%
\usepackage{dcolumn}
\usepackage{bm}
\usepackage{color}

\usepackage{graphicx}
\usepackage{dcolumn}
\usepackage{bm}
\usepackage{amsmath}
\usepackage{tabu}
\usepackage{multirow}
\usepackage[latin1]{inputenc}
\usepackage{subfigure}
\usepackage{braket}
\usepackage{soul}
\setstcolor{red}
\usepackage{xcolor}
\usepackage{verbatim}
\usepackage[dvipdfm, pdfstartview=FitH, CJKbookmarks=true, bookmarksnumbered=true,
bookmarksopen=true, colorlinks, pdfborder=001, linkcolor=blue, anchorcolor=blue, citecolor=blue]{hyperref}
\newcommand{\sr}[1]{\textcolor{black}{#1}}

\begin{document}
\preprint{Regular article}



\title{Impurity-induced bound states inside the superconducting gap of FeSe}

\author{Lin Jiao}
\affiliation{Max Planck Institute for Chemical Physics of Solids,
N\"othnitzer Stra\ss e 40, 01187 Dresden, Germany}
%
%
\author{Sahana R\"o{\ss}ler}
\email{roessler@cpfs.mpg.de}
\affiliation{Max Planck Institute for Chemical Physics of Solids,
N\"othnitzer Stra\ss e 40, 01187 Dresden, Germany}
\author{Cevriye~Koz}
\affiliation{Max Planck Institute for Chemical Physics of Solids,
N\"othnitzer Stra\ss e 40, 01187 Dresden, Germany}
%
%
%
%
%
\author{Ulrich Schwarz}
\affiliation{Max Planck Institute for Chemical Physics of Solids,
N\"othnitzer Stra\ss e 40, 01187 Dresden, Germany}
\author{Deepa Kasinathan}
\affiliation{Max Planck Institute for Chemical Physics of Solids,
N\"othnitzer Stra\ss e 40, 01187 Dresden, Germany}
\author{Ulrich K. R\"o{\ss}ler}
\affiliation{IFW Dresden, Postfach 270016, 01171 Dresden, Germany}
\author{Steffen Wirth}
\email{wirth@cpfs.mpg.de}
\affiliation{Max Planck Institute for Chemical Physics of Solids,
N\"othnitzer Stra\ss e 40, 01187 Dresden, Germany}

\date{\today}

\begin{abstract}

We investigate the local density of states in the vicinity of a native dumbbell defect arising from an Fe vacancy in FeSe single crystals. The tunneling spectra close to the impurity display two bound states inside the superconducting gap, equally spaced with respect to zero energy but asymmetric in amplitude. \sr{Using spin-polarized density functional theory (DFT) calculations on realistic slab models with Fe vacancy, we show that such a defect does not induce a local magnetic moment. Therefore, the dumbbell defect is considered as non-magnetic. Thus, the in-gap bound states emerging from a non-magnetic defect-induced pair-breaking suggest a sign changing pairing state in this material.}

\end{abstract}


\maketitle
\section{Introduction}
The Fe-based superconductors (Fe-SC) are currently in vogue \cite{Si2016, Hir2016} because of their high superconducting transition temperatures ($T_c$), which stand second only to the high-$T_c$ cuprates. Unlike the cuprates, Fe-SCs are semimetallic materials in which multiple orbitals are involved in superconductivity \cite{Leb2007,Sin2008} and Hund's rule couplings are considered as responsible for intermediate electron correlations \cite{Hau2009,Qaz2009,Geo2013}. In spite of intensive research, the superconducting gap symmetry and structure of Fe-SC have proven to be most challenging to unveil owing to multiple bands crossing the Fermi energy \cite{Hir2011,Hir2016}. 
A spin-fluctuation mediated inter-band scattering between the hole and electron pockets is considered as a plausible pairing mechanism for superconductivity in these materials. This mechanism leads to a sign-changing $s^{\pm}$ symmetry for the superconducting order parameter (OP) with wave-functions of different phases on the hole and electron Fermi pockets \cite{Maz2008}. Alternative theories consider a conventional sign-preserving $s^{++}$ symmetry induced by orbital fluctuations originating from the phonon mediated electron-electron interaction \cite{Kon2010}. 

However, the structurally simplest member FeSe and its related compounds at first glance appear to pose a serious challenge to the spin-fluctuation-based theories. This is because, unlike several parent compounds of the Fe-pnictides, FeSe does not display long-range magnetic order at ambient pressure. Instead, it undergoes an enigmatic structural (also called nematic) transition at $T_s\approx$ 87~K at which the $C_4$-symmetry of both the lattice \cite{Mc2009a} as well as the underlying electronic structure is broken \cite{Song2011}. The origin of nematicity and its relationship to superconductivity are both highly debated --- spin, orbital and even charge fluctuations are considered as likely candidates \cite{Fer2014}. Angle-resolved photoemission spectroscopy (ARPES) and quantum oscillation experiments detect one hole Fermi pocket at the center of the Brillouin zone, but allude to the presence of two electron pockets at the zone boundary \cite{Mal2014,Wat2015}. Several experiments detected at least two superconducting gaps \cite{Dong2009,Kha2010,Lin2011,Abd2013,Kas2014,Bour2016,Tek2016,Li2016,Lin2016} with strong anisotropy at least for one of the gaps \cite{Lin2011,Kas2014,Bour2016,Tek2016,Li2016,Lin2016}. Whether the anisotropic gap actually contains accidental nodes or only deep minima is a question of debate \cite{Kas2014,Bour2016,Tek2016,Li2016,Lin2016}.  

One of the powerful experiments to investigate the structure of the gap is to conduct scanning tunneling microscopy and spectroscopy (STM/S) measurements in the vicinity of non-magnetic impurities \cite{Bal2006}. The tunneling conductance $g(V,r)$ = d$I(V,r)$/d$V$ (where $I$ is the current and $V$ the applied voltage) is, with simplifying approximations, directly proportional to the local density of states (LDOS). As a hallmark of $s^{\pm}$ symmetry, both non-magnetic and magnetic impurities are expected to produce bound states inside the superconducting gap \cite{Bea2012,Wang2013,Miz2014,Hir2015,Akb2010}, whereas for $s^{++}$ symmetry such states can be induced only by magnetic scatterers owing to the time reversal symmetry breaking. The energy of the bound states with respect to the maximum of the superconducting energy gap and its spatial distribution both contain information about the OP symmetry \cite{Bea2012}. 

Experimentally, impurity-induced in-gap bound states unequally spaced with respect to the Fermi energy have been previously observed in FeSe thin films close to a single Fe adatom or at a site of a single Se vacancy \cite{Song2011}. For single layer FeSe film on a SrTiO$_{3}$ substrate, where the hole band is absent at the center of the Brillouin zone, in-gap bound states have been found only in the vicinity of magnetic impurities \cite{Fan2015}. In the case of single crystals, a single sharp resonance peak outside \cite{Kas2014} as well as inside \cite{Moo2015} the superconducting gap has been reported. \sr{Although theories predict two in-gap features at positive and negative bias voltages \cite{Hir2016,Bang2017}, these features tend to have different weights \cite{Hir2016} and are up to now observed only at either positive or negative bias.} In fact, unlike in a one-band superconductor, the formation of an impurity bound state inside the superconducting gap of a multiband system is more complicated, and requires a fine tuning of interband and intraband scattering potentials \cite{Hir2011,Hir2015}. Further, the superconducting gap in bulk FeSe is only of the order of 2 meV. Hence, a high energy resolution and low temperatures are required to resolve the bound states inside the superconducting gap. Here we present an observation of two in-gap bound states in the tunneling spectra taken on FeSe single crystals close to an apparent selenium (Se) dumbbell defect \sr{induced by an Fe vacancy \cite{Den2016}. Further, utilizing spin-polarized density functional theory (DFT) calculations on
realistic slab models with a specific Fe vacancy, we show that the moments on the neighboring Fe atoms remain very small and itinerant, thereby suggesting that the defect site remains invariant to the time reversal symmetry.} 


\begin{figure}[t!]
\centering
	    \includegraphics[clip,width=0.95\columnwidth]{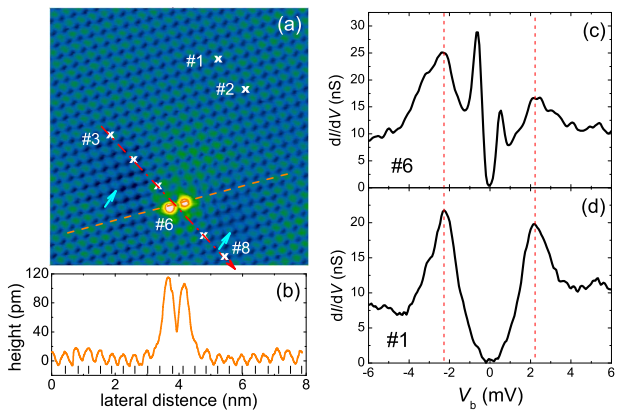}
	\caption{(a) A topography of FeSe on an area of 8 $\times$ 8 nm$^2$ containing a Se-dumbbell impurity. The tunneling spectra measured along the red arrow in (a) are presented in Fig. 2. The cyan arrows indicate the position of the electronic dimers. (b) A line scan along the dashed orange line in (a) displaying the height of the dumbbell. (c) Tunneling spectrum measured close to the dumbbell defect (at \#6) displaying in-gap bound states (d) Tunneling spectrum measured away (at \#1) from the dumbbell defect displaying the clean superconducting gap. The vertical lines in (c) and (d) identify the coherence peaks of the superconducting gap.}
\end{figure}
\begin{figure}[t!]
\centering
	    \includegraphics[clip,width=0.95\columnwidth]{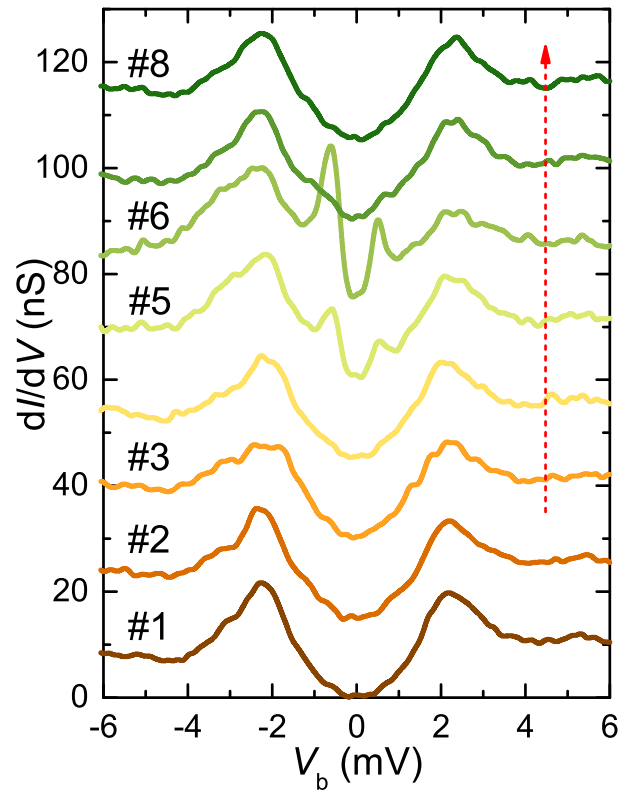}
	\caption{Tunneling spectra taken at positions marked in Fig.~1(a). The red arrow indicates the spectra taken along the red arrow in Fig.~1(a). Spectra $\#$5 (close to the dumbbell) and $\#$6 (on top of the dumbbell) display in-gap bound states. Note that the superconducting gap magnitude remains unaffected by the impurity potential.}
\end{figure}

\section{Experimental}
The single crystals used for the measurements were grown by a chemical vapor transport method \cite{Koz2014,Ros2016}. \sr{The crystals displayed a superconducting transition temperature $T_c \approx 8.5$~K \cite{Lin2016}.}
The samples were cleaved $in~situ$ at 20~K in an ultrahigh vacuum chamber. STM/S experiments were conducted at 0.35 K. The topography was measured at a current set point $I_{\mathrm{sp}}$ =~100 pA and a bias voltage $V_\mathrm{b}$ =~10 mV. The tunneling spectra were obtained using a lock-in technique with a modulation voltage $V_{\mathrm{mod}}$ =~0.05 m$V_{\mathrm{rms}}$. The energy resolution of the spectroscopic measurement is about 0.16 meV. The DFT calculations were performed
using the full-potential local orbital (FPLO) approach \cite{Kla1999}, and the generalized gradient approximation (GGA) \cite{Per1996}. 

\begin{figure*}[t!]
\centering
	    \includegraphics[clip,width=1.6\columnwidth]{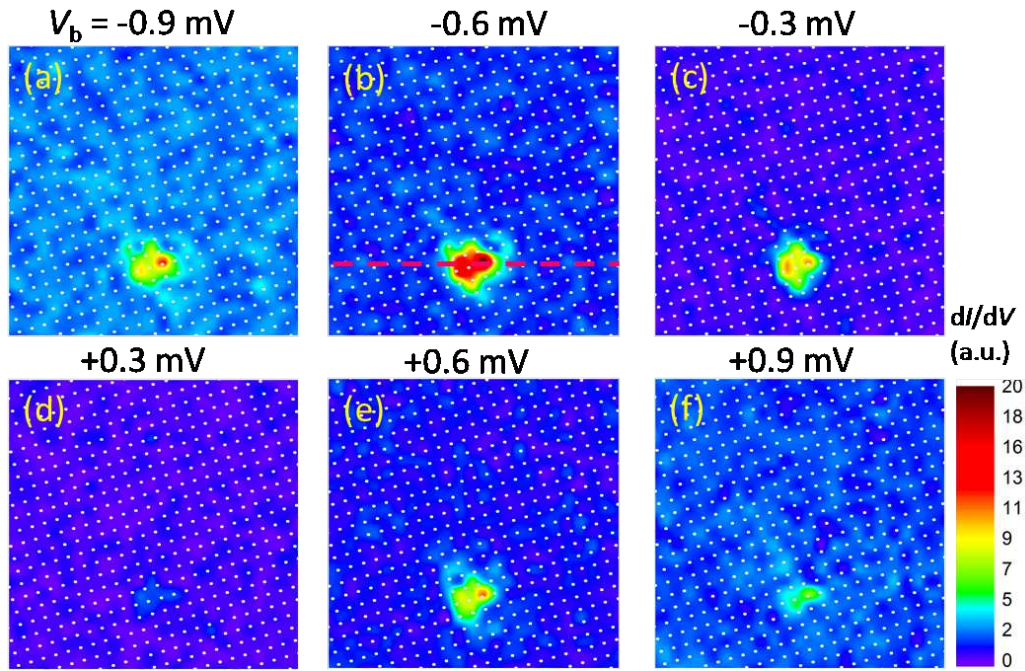}
	\caption{(a)-(f) Conductance maps $g(V,r)$ at selected bias voltages $V_b$ taken on the topography in Fig.~1(a) displaying the spatial distribution of the in-gap bound states. The white spots represent the position of the surface Se lattice. At the bound state peak maximum  $|V_b|$=0.6 mV, the real space length scale of the in-gap bound state extends up to four Se-lattice units. The data were measured with a resolution of 0.32 nm/pixel. A second order interpolation was used to generate the images. The tunneling amplitude measured along the red line in (b) are plotted in Fig. 4.}
\end{figure*}

\section{Results and Discussion}
As a result of the crystal structure with van der Waals bonds between adjacent Se layers, the cleaving exposes a Se-terminated surface as shown in Fig.~1(a). The Se-Se distance $a_\mathrm{Se-Se} = 3.7(1)$ \AA~observed here is consistent with the distance of 3.7702(5)~\AA~found by x-ray diffraction on our crystals \cite{Koz2014}. In the topographic image, two bright atoms (dumbbell) of an apparent height of about 100--110~pm (Fig.~1(b)) are visible. 
This type of impurities is ubiquitous on both, cleaved surfaces of single crystals \cite{Kas2014,Wata2015,Lin2016} as well as on the high-quality thin film surfaces grown by molecular beam epitaxy (MBE)  \cite{Song2012,Den2016}. Further, these impurities induce small unidirectional depressions in the density of states, known as ``electronic dimers'' \cite{Song2012}, marked by the cyan arrows in Fig.~1(a). While the extent of the dumbbell is close to $a_\mathrm{{Se-Se}}$, the length scale of the electronic dimers amounts to $\approx~8\sqrt{2}a_\mathrm{Se-Se}$ = 16~$a_\mathrm{Fe-Fe}$. \sr{The $C_{2}$ symmetry of the electronic dimers indicates an anisotropic scattering mechanism, which we previously reported based on transport measurements \cite{Ros2015}.} Although the dumbbell consists of Se atoms, the position of the actual defect is most likely at the Fe-site, situated underneath the center of the dumbbell. In the case of MBE-grown thin films, the number of dumbbells increased substantially if the films were grown in a Fe-deficient atmosphere, which suggested that the dumbbells correspond to Fe-vacancies \cite{Den2016}. In addition, a DFT based calculation suggested that Fe-site vacancies can perturb orbitals on neighboring Se-sites, thereby producing atomic dumbbells \cite{Den2016}. Alternatively, solutions of the Bogoliubov-de Gennes equations with a first-principles Wannier analysis reproduces the atomic dumbbells by putting a simple non-magnetic impurity at an Fe-site \cite{Cho2014}. Thus it can be concluded that the dumbbell defects are stemming from either a vacancy or a nonmagnetic impurity at the Fe-site. \sr{However, these calculations did not consider spin polarization, which is essential for determining a possible defect-induced local magnetic moment.} 

\sr{To verify theoretically if there is a defect-related change of the magnetic state in FeSe, we constructed $c$-axis oriented $2 \times 3 \times 2$ slab with Se-termination. Two types of defects, one with a single Se vacancy at the surface and the other with a single Fe vacancy below the Se-covered surface were considered, as shown in the Appendix, Fig. A1.
As mentioned above, the latter Fe vacancy defect was previously identified to explain dumbbell defects in an STM study supported by a DFT calculation on a single FeSe-layer slab, yet not considering spin-polarization effects \cite{Den2016}. The magnetism in bulk FeSe crystals itself is suppressed owing to strong spin-fluctuations which prevent stabilization of long-range magnetic order \cite{Ima2009,Wang2015}. In contrast to the experimental results, bare DFT on FeSe shows a strongly spin-polarized antiferromagnetic ground-state \cite{Sub2008}. The effect of the large-scale spin-fluctuations are considered by the rescaling of the exchange-correlation (xc) potential through a reduction factor $s~<~1$. This scheme was proposed by Ortenzi $et~al.$ \cite{Ort2012} as modified Stoner theory. In DFT calculations it has been successfully used to model electronic and spin-structure of Fe-based superconducting materials \cite{Ort2015}, including FeSe \cite{Lis2015,Ess2016}. For bulk FeSe we find the spin-polarization to vanish in the GGA calculations with  $s~<~0.70$ (Fig. A2, Appendix) in agreement with calculations in Ref. \onlinecite{Ess2016}. For the slabs with defects we performed calculations for a range of xc-reduction factors. As a general conclusion of these studies, we do
find only small modifications of the spin-polarization of clean slabs. The symmetry breaking of defects is able to induce small spin polarization spread over many atoms in both Fe-layers
of the slab with concomitant very-small and mainly antiparallel spin polarization on Se-atoms. Most significantly, calculations for  $s~=~ 0.65$, $i.~e.,$ close to the threshold for acquiring
a spin-polarization in the bulk, in the vicinity of the Se-vacancy, neighboring Fe atoms develop a moment of 1.59  $\mu_\mathrm{B}$/Fe, see Appendix, Table AI. Similar observations have been made in Fe-pnictides with arsenic vacancy \cite{Gri2011,Kik2015}. Such a large moment might induce blocked paramagnetic units depending on the time scale and the temperature. On the other hand, the moments in direct vicinity of the Fe vacancy were found to be in the range 0.1 - 0.3 $\mu_\mathrm{B}$/Fe (see Appendix, Table AII). Thus, the observed spin polarization has a metallic-like itinerant character and no stable localized spin-state was observed. From these calculations, it is fair to conclude that the Fe vacancy-induced dumbbell defects most likely are not associated with a significant breaking of the time reversal symmetry.}

\begin{figure}[t!]
\centering
	    \includegraphics[clip,width=0.9\columnwidth]{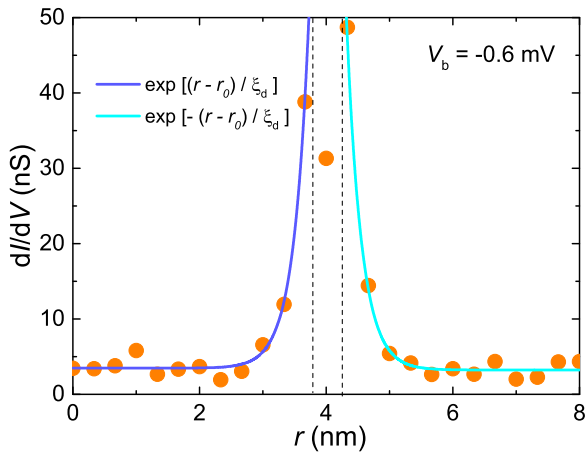}
	\caption{The data points correspond to the tunneling amplitude measured along the red line in Fig.~3(b). The solid lines are obtained from an exponential fit to the data. The parameter $\xi_\mathrm{d}$ represents the characteristic length scale of the decay and $r_{0}$ is the position of the impurity.} 
\end{figure}

\sr{Thus, any signature of pair-breaking, if observed in the tunneling spectra in response to the non-magnetic dumbbell defects, is expected to point to a sign-changing pairing mechanism in FeSe.} In order to verify this, we performed spatially resolved conductance maps $g(V,r)$ on the topography displayed in Fig.~1(a).
A tunneling spectrum d$I(V,r)$/d$V$ close to the dumbbell defect (at the point indicated by \#6 in Fig.~1(a)) is presented in Fig.~1(c), where strong resonance peaks at $V_b \approx~ \pm$0.6 mV can be seen. 
The intensities of the bound state peaks are asymmetric, which is consistent with the underlying asymmetry in the DOS of the hole and electron bands \cite{Ros2015}. For comparison, a d$I(V,r)$/d$V$ curve measured away from the dumbbell defect (at the point indicated by \#1 in Fig.~1(a)) is shown in Fig.~1(d). \sr{Further}, the tunneling spectra collected along the arrow indicated in Fig.~1(a) are presented in Fig.~2. All spectra, including those close to the dumbbell, displayed a superconducting gap with coherence peaks at bias voltages $|V_b|$ = 2.25(5) meV. 
\sr{The appearance of coherence peaks at the defect site suggests that either it is a result of spatial superimposition of the tunneling response at the pair-breaking impurity site and the superconducting bulk or it is due to a partial suppression of superconductivity at the impurity site. In the light of a recently proposed orbital-selective Cooper pairing mechanism \cite{Spr2016}, it is reasonable to assume that the impurity scatterings cause pair-breaking only in those orbitals where the pairing is weak, $i.e.$, in a region close to the superconducting gap-minimum in $k$-space.}

In Fig.~3(a-f), the $g(V,r)$ maps at selected bias voltages $V_b$ taken on the topography in Fig.~1(a) are presented. Far away from the dumbbell defects, the smaller superconducting gap is homogeneous as can be inferred from Fig.~3(b-e). Some spatial variation was found in the tunneling amplitude for bias voltages $|V_b|$ = 0.9 mV, which can be seen in Fig.~3(a) and (f).
This is due to a steep increase in $g(V,r)$ close to the edge of the gap minimum.
In Figs. 3(b) and (e) we present the spatially resolved  maps at $V_b$ = $\pm$ 0.6 mV, i.e. at the maximum of the resonance for the bound states. These maps visualize the rather localized nature of the resonance. \sr{Such a localization indicates that the bound states correspond to broken Cooper pairs created from the quasiparticle scattering by the impurity potential and are not related to the small superconducting gap at comparable energy reported previously \cite{Lin2016} $i.e.$, the bound states and the small gap likely occur at different parts of the $k$-space. Furthermore, the localization also suggests that the bound states are decoupled from the extended superconducting states (see following paragraph) and hence, they likely exist just below the gap edge of the large, anisotropic gap with a gap minimum of 0.88(2) meV \cite{Lin2016}. Similar behavior has also been reported for LiFeAs \cite{Chi2016}}. 



In order to obtain a quantitative estimate of the spatial extent of the bound states, we plotted in  Fig.~4 the amplitude as a function of distance along the line drawn in Fig.~3(b). The two vertical lines in Fig.~4 represent the positions of the two dumbbell constituents. On both sides, the amplitude decays exponentially as $\mathrm{exp}{(|r-r_0|/\xi_\mathrm{d})}$ \cite{Yin2015}. Here, $r$ is the distance, $r_0$ is the position of the impurity, and $\xi_\mathrm{d}$ represents the spatial decay length. We obtain $2\xi_\mathrm{d} \approx$ 5 \AA,~\sr{which is approximately an order of magnitude smaller than the superconducting coherence length $\xi$ $\approx 40$ \AA~in FeSe \cite{Hsu2008}.} Both the exponential decay of the LDOS as well as such a small value of $\xi_\mathrm{d}$ further confirm the localization of the impurity states. The value obtained here is slightly lower than the reported values of $\xi_\mathrm{d}$ in Fe$_{1+x}$(Se,Te) \cite{Yin2015}. 

The asymmetric in-gap bound states observed here are in agreement with those obtained from an on-site LDOS calculation applied to a four-band model with moderate inter-band scatterings for the case of a nodeless anisotropic $s^{\pm}$ superconductor \cite{Bea2012}. Further, our conclusions presented here are in excellent agreement with the experimental study reported very recently \cite{Spr2016} in which the Bogoliubov quasipaticle scattering interference (BQPI) was used to determine the superconducting gap symmetry as nodeless with an OP changing sign between the hole and electron pockets.
\section{conclusions}
In summary, using scanning tunneling microscopy/spectroscopy, we observed in-gap bound states in the vicinity of Fe vacancy-induced dumbbell defects. Utilizing spin polarized DFT, we show that this type of defect does not develop a local magnetic moment. Since the observation of a localized resonance near a non-magnetic impurity is one of the clearest indications of a $s^{\pm}$ symmetry of the order parameter \cite{Hir2016}, our experimental results are in favor of theories which propose a sign-changing structure of the superconducting gap in FeSe.

\section{Appendix}
In the Appendix, we present results of spin-polarized density functional theory (DFT) calculations for the defect slabs of FeSe.  We constructed $c$-axis oriented $2 \times 3 \times 2$ slab with Se-termination. Two types of defects, one with a single Se vacancy at the surface and the other with a single Fe vacancy below the Se-covered surface were considered. The results show a significant magnetic moment close to Se vacancy. However, in the vicinity to Fe vacancy, the observed spin polarization has a metallic-like itinerant character and no stable localized spin-state was observed. From these calculations, we conclude that the Fe vacancy-induced dumbbell defects found in FeSe can be regarded as non-magnetic impurity which are not associated with a significant breaking of the time reversal symmetry.  

\begin{figure}[b!]
\includegraphics[clip,width=6.9cm]{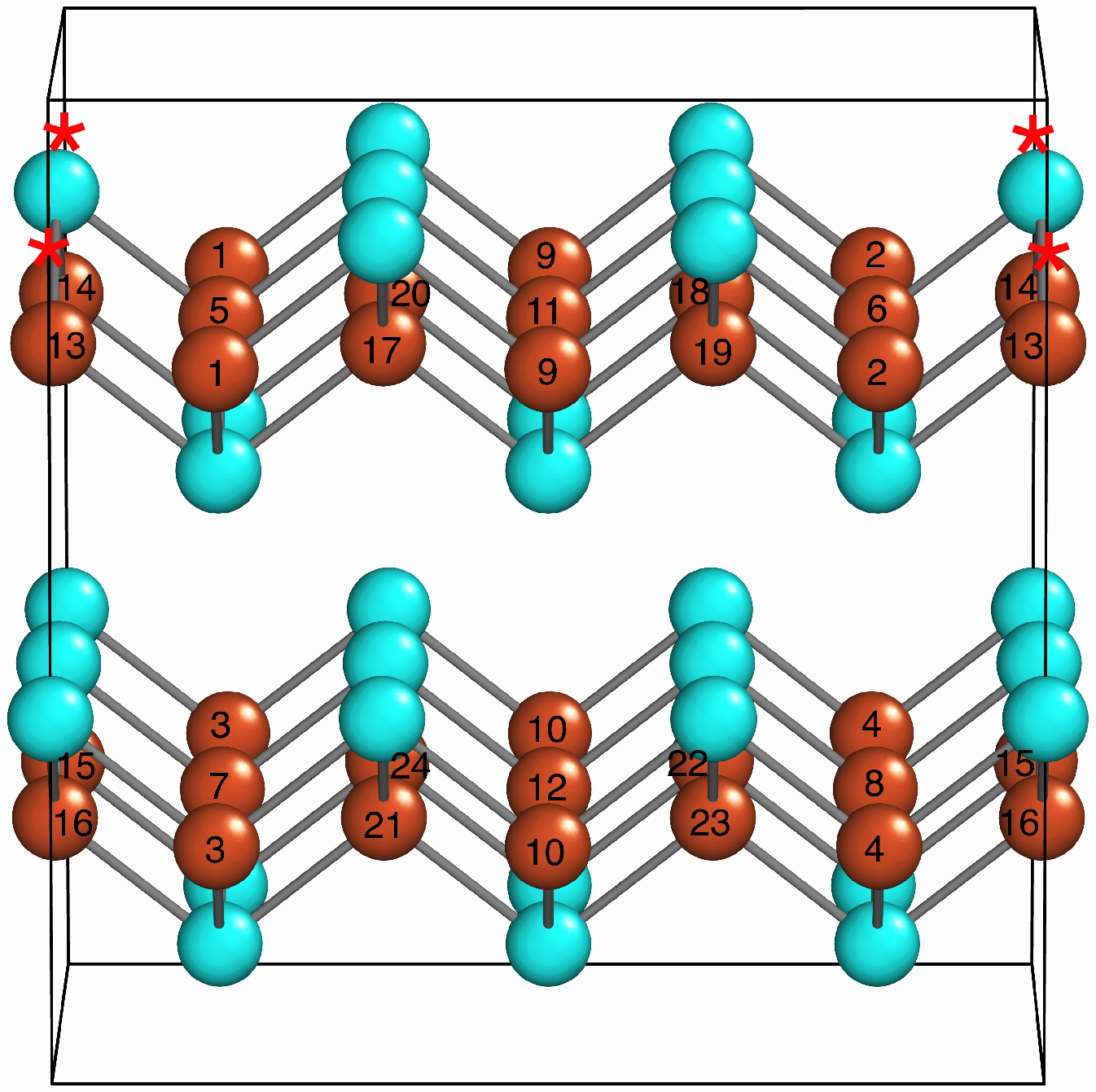}
\hspace{1cm}
\includegraphics[clip,width=6.9cm]{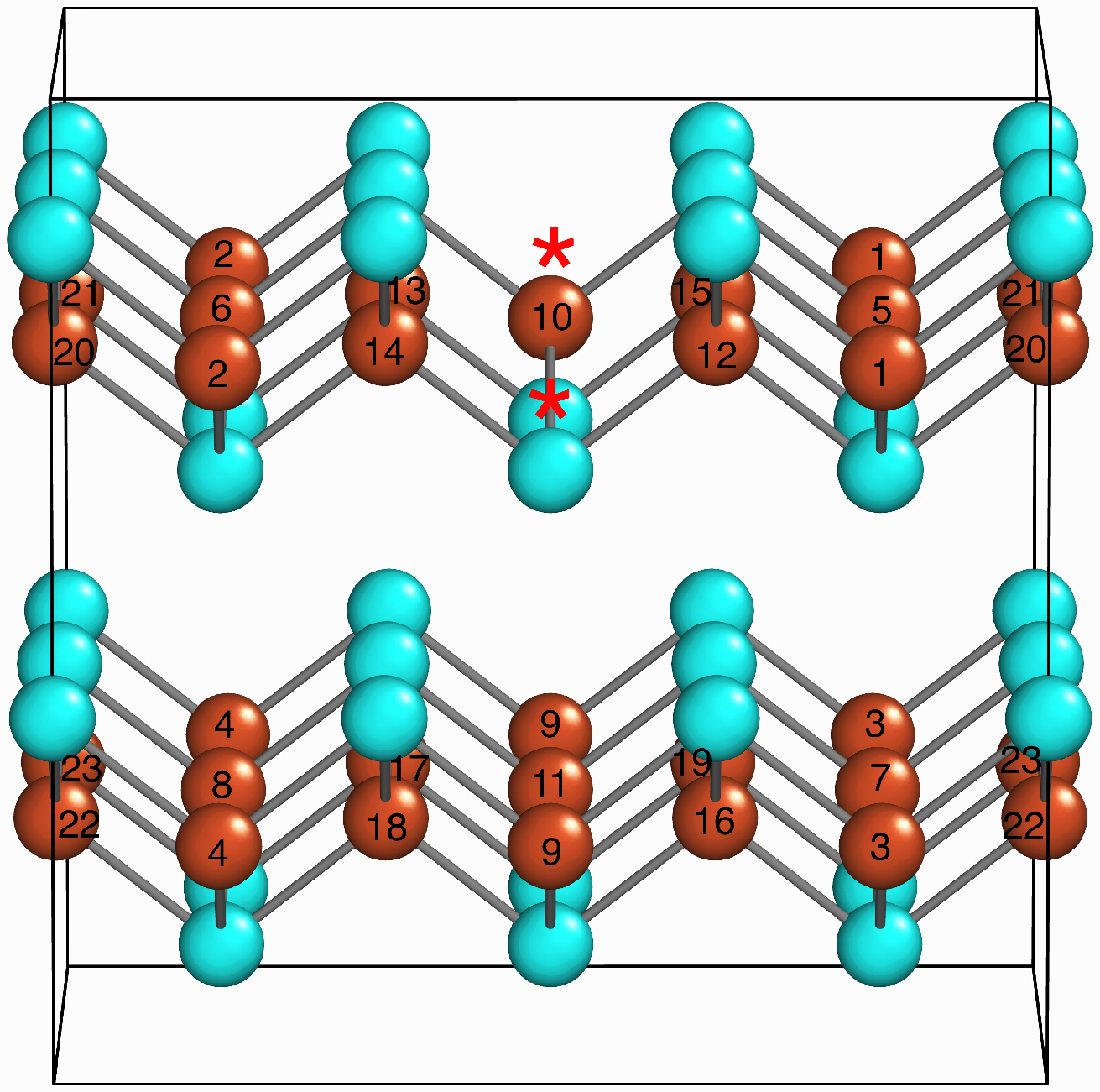}
\setcounter{figure}{0}
\renewcommand{\figurename}{FIG. A$\!\!$}
\caption{The crystal structure of the FeSe slabs with Se-defect (top) and Fe-defect (bottom). The position of the defects
are marked by the star symbol. The numbered spheres represent the Fe atoms while the non-numbered ones represent
the Se atoms. The numbering is used to document the evolution of the Fe spin-moment as a function of $s$, listed
in Tables~SI and SII.
}
\end{figure}
\begin{figure}[t]
\centering
	    \includegraphics[clip,width=0.9\columnwidth]{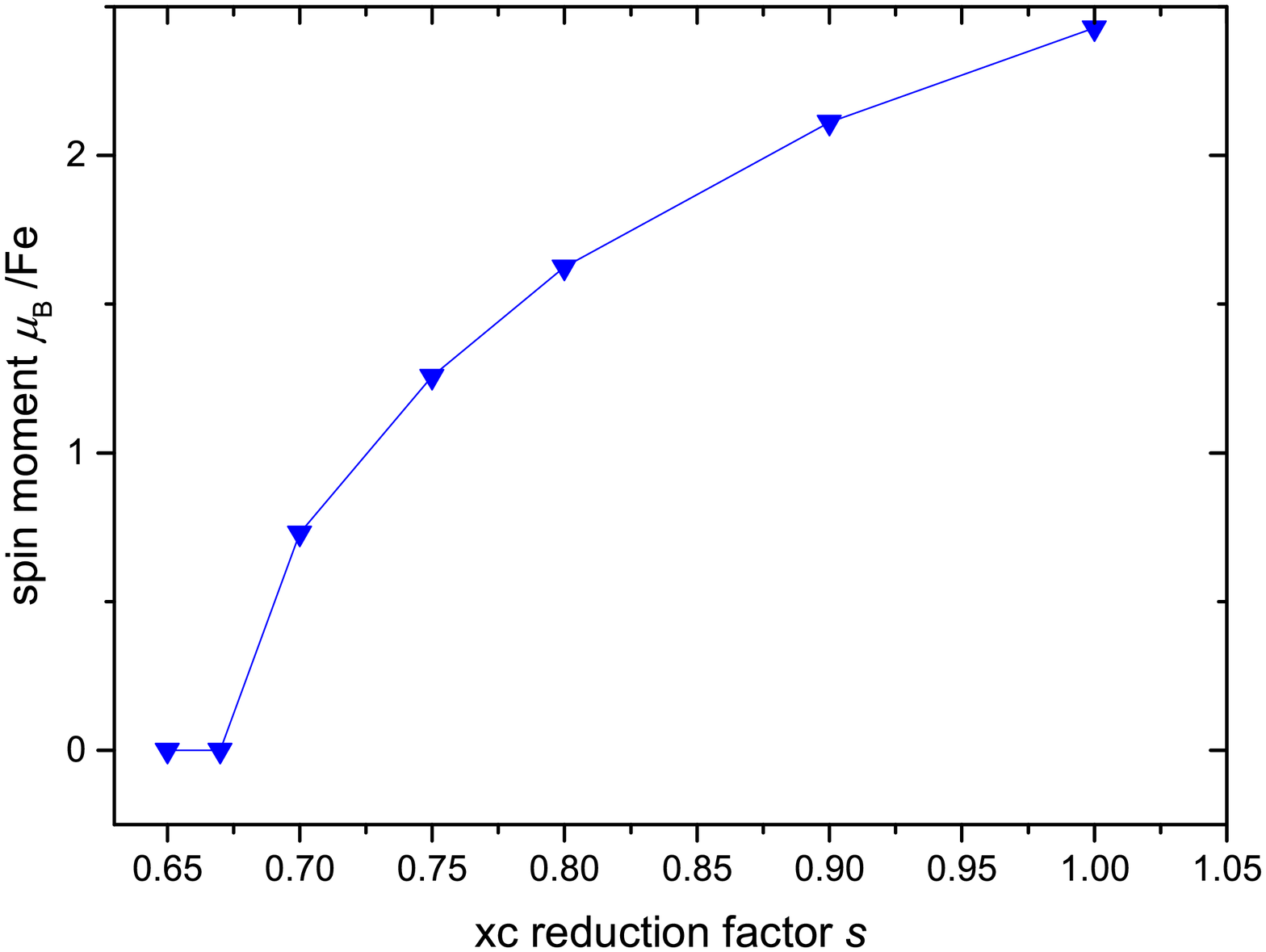}
			\setcounter{figure}{1}
			\renewcommand{\figurename}{FIG. A$\!\!$}
	\caption{Fe-moment in bulk tetragonal FeSe, from spin-polarized DFT-GGA calculations
with rescaled exchange-correlations $s$.} 
\end{figure}

The density functional theory (DFT) calculations have been performed using the FPLO code (https://www.fplo.de/, version 14.00).
In order to keep the computational efforts manageable
a $c$-axis oriented slab of $2 \times 3 \times 2$ unit cells with Se-termination is used (Fig. A1),
where the atomic position have been fixed as in the ideal tetragonal bulk-crystal. 
It is not expected that this simplification causes strong effects regarding
the propensity of a defected FeSe-surface to acquire sizable localized magnetic states.
As the DFT-approach is not very precise regarding bonding properties,
a geometric optimization of the slab is not meaningful,
in particular, it would be unable to model the van-der-Waals-bonding.
However, we have checked that forces on atoms in these idealized 
positions are relatively weak. 

%
To act as a metric for calibrating the range of reduction factors ($s$) for
the exchange correlation (xc) potential, we first calculate the magnetic moments 
of the bulk with tetragonal structure
and ferromagnetic spin-polarization.
Collected in Fig. A2, is the change in the Fe spin-moment as a function of  
$s$ which shows a complete suppression of the spin-moment for $s$ = 0.675. 
Next, we calculate the Fe spin-moments for two slabs, one containing a Se vacancy and
the other a Fe vacancy. For clarity, all the Fe sites in the slab are 
numbered. The Se vacancy is located on top of the $\#$13 to $\#$14 Fe site, while the Fe vacancy is set close to $\#$10 Fe site in Fig. A1, top and bottom panels, respectively. Taking into consideration the results from the bulk FeSe calculations, 
the exchange-correlations for the two different defect slabs are rescaled
from $s$ = 1 to 0.60 and the obtained Fe spin-moments are tabulated in 
Tables~AI and AII.
Considering the Fe sites that are far away from the defects, both in the Se and Fe defected slabs,
the reduction in $s$ results in a
loss of the Fe spin-moment in the same vein as reported above for the bulk. 
On the other hand, the Fe sites close to the Se-defects retain a large spin-moment for $s$ up to 0.60,
while the Fe sites close to the Fe-defects behave like the bulk and also loose their spin moments.

\begin{table*}[b]
\renewcommand{\tablename}{TABLE. A$\!\!$}
   \caption{The spin moments of Fe in an Se-defected slab of FeSe as a function of rescaled exchange-correlations $s$. The site numbers in this table correspond to the Fe-sites denoted in Fig.~A1, top panel. The spin moments of the Fe atoms in the vicinity of the defect for different values of $s$ are highlighted in boldface.}
		\vspace{0.25cm}
  \begin{tabular}{|c|c|c|c|c|c|c|c|c|c|}
  \hline
 site  & $s$ = 1  & $s$ = 0.95  &  $s$ =0.90  & $s$ = 0.85 & $s$ = 0.80 & $s$ = 0.75 & $s$ = 0.70 & $s$ = 0.65 & $s$ = 0.60   \\
 \hline
\textbf{1}   &   \textbf{2.84}   &     \textbf{2.59}   &    \textbf{2.50}   &   \textbf{2.49}   &    \textbf{2.56}   &    \textbf{2.43}  &    \textbf{1.82}   &   \textbf{1.58}   &   \textbf{1.30}   \\
\textbf{2}   &   \textbf{2.84}   &     \textbf{2.59}   &    \textbf{2.50}   &   \textbf{2.49}   &    \textbf{2.56}   &    \textbf{2.43}  &    \textbf{1.82}   &   \textbf{1.58}   &   \textbf{1.30}   \\
 3   &   2.29   &     1.80   &    1.82   &   1.89   &    2.01   &    1.80  &    0.02   &   0.00   &   0.00   \\
 4   &   2.29   &     1.80   &    1.82   &   1.89   &    2.01   &    1.80  &    0.02   &   0.00   &   0.00   \\
 5   &   2.52   &     2.06   &    1.92   &   2.07   &    2.20   &    1.96  &    1.06   &   0.73   &   0.37   \\
 6   &   2.52   &     2.06   &    1.92   &   2.07   &    2.20   &    1.96  &    1.06   &   0.73   &   0.37   \\
 7   &   2.28   &     1.76   &    1.81   &   1.87   &    1.99   &    1.79  &    0.03   &   0.01   &   0.00   \\
 8   &   2.28   &     1.76   &    1.81   &   1.87   &    1.99   &    1.79  &    0.03   &   0.01   &   0.00   \\
 9   &   2.55   &     2.08   &    2.05   &   2.14   &    2.27   &    2.05  &    0.43   &   0.43   &   0.32   \\
10   &   2.31   &     1.81   &    1.83   &   1.89   &    2.03   &    1.81  &    0.05   &   0.01   &   0.00   \\
11   &   2.46   &     1.97   &    1.94   &   2.06   &    2.20   &    1.96  &    1.13   &   0.74   &   0.26   \\
12   &   2.32   &     1.82   &    1.87   &   1.90   &    2.04   &    1.83  &    0.06   &   0.02   &   0.00   \\
\textbf{13}   &   \textbf{2.87}   &     \textbf{2.64}   &    \textbf{2.53}   &   \textbf{2.55}   &    \textbf{2.58}   &    \textbf{2.46}  &    \textbf{1.85}   &   \textbf{1.59}   &   \textbf{1.28}   \\
\textbf{14}   &   \textbf{2.87}   &     \textbf{2.64}   &    \textbf{2.53}   &   \textbf{2.55}   &    \textbf{2.58}   &    \textbf{2.46}  &    \textbf{1.85}   &   \textbf{1.59}   &   \textbf{1.28}   \\
15   &   2.28   &     1.77   &    1.81   &   1.89   &    1.99   &    1.80  &    0.01   &   0.00   &   0.00   \\
16   &   2.28   &     1.77   &    1.81   &   1.89   &    1.99   &    1.80  &    0.01   &   0.00   &   0.00   \\
17   &   2.45   &     1.98   &    1.92   &   2.03   &    2.16   &    1.92  &    0.54   &   0.31   &   0.14   \\
18   &   2.45   &     1.98   &    1.92   &   2.03   &    2.16   &    1.92  &    0.54   &   0.31   &   0.14   \\
19   &   2.45   &     1.98   &    1.92   &   2.03   &    2.16   &    1.92  &    0.54   &   0.31   &   0.14   \\
20   &   2.45   &     1.98   &    1.92   &   2.03   &    2.16   &    1.92  &    0.54   &   0.31   &   0.14   \\
21   &   2.31   &     1.80   &    1.84   &   1.89   &    2.02   &    1.81  &    0.01   &   0.00   &   0.00   \\
22   &   2.31   &     1.80   &    1.84   &   1.89   &    2.02   &    1.81  &    0.01   &   0.00   &   0.00   \\
23   &   2.31   &     1.80   &    1.84   &   1.89   &    2.02   &    1.81  &    0.01   &   0.00   &   0.00   \\
24   &   2.31   &     1.80  &     1.84  &    1.89  &     2.02   &    1.81 &     0.01  &    0.00   &   0.00   \\  
   \hline
   \end{tabular}
\end{table*}  
\begin{table*}[b]
\renewcommand{\tablename}{TABLE. A$\!\!$}
   \caption{The spin moments of Fe in an Fe-defected slab of FeSe as a function of rescaled exchange-correlations $s$. The site numbers in this table correspond to the Fe-sites denoted in Fig.~A1, bottom panel. The spin moment of the Fe atom in the vicinity of the defect for different values of $s$ is highlighted in boldface.}
	\vspace{0.25cm}
  \begin{tabular}{|c|c|c|c|c|c|c|c|c|c|}
  \hline
 site  & $s$ = 1  & $s$ = 0.95  &  $s$ =0.90  & $s$ = 0.85 & $s$ = 0.80 & $s$ = 0.75 & $s$ = 0.70 & $s$ = 0.65 & $s$ = 0.60   \\
 \hline
  1 & 1.69  &   1.52  &  1.39  &  1.38  &  1.37   &  1.40    &  0.14    &  0.16   &   0.01 \\   
 2 & 1.69  &   1.52  &  1.39  &  1.38  &  1.37   &  1.40    &  0.14    &  0.16   &   0.01 \\    
 3 & 2.14  &   1.71  &  1.75  &  1.84  &  1.82   &  1.85    &  0.12    &  0.10   &   0.04 \\   
 4 & 2.14  &   1.71  &  1.75  &  1.84  &  1.82   &  1.85    &  0.12    &  0.10   &   0.04 \\   
 5 & 2.11  &   1.84  &  1.67  &  1.65  &  1.68   &  1.66    &  0.24    &  0.07   &   0.01 \\   
 6 & 2.11  &   1.84  &  1.67  &  1.65  &  1.68   &  1.66    &  0.24    &  0.07   &   0.01 \\   
 7 & 2.13  &   1.72  &  1.76  &  1.82  &  1.76   &  1.83    &  0.11    &  0.02   &   0.03 \\   
 8 & 2.13  &   1.72  &  1.76  &  1.82  &  1.76   &  1.83    &  0.11    &  0.02   &   0.03 \\   
 9 & 2.23  &   1.67  &  1.64  &  1.72  &  1.70   &  1.74    &  0.12    &  0.01   &   0.04 \\   
\textbf{10} & \textbf{1.84}  &   \textbf{1.49}  &  \textbf{1.11}  &  \textbf{1.44}  &  \textbf{1.48}   &  \textbf{1.47}    &  \textbf{0.54}    &  \textbf{0.30}   &   \textbf{0.06} \\   
11 & 2.23  &   1.74  &  1.75  &  1.91  &  1.81   &  1.92    &  0.19    &  0.18   &   0.05 \\   
12 & 2.38  &   1.70  &  1.59  &  1.76  &  1.78   &  1.77    &  0.36    &  0.10   &   0.01 \\   
13 & 2.38  &   1.70  &  1.59  &  1.76  &  1.78   &  1.77    &  0.36    &  0.10   &   0.01 \\   
14 & 2.38  &   1.70  &  1.59  &  1.76  &  1.78   &  1.77    &  0.36    &  0.10   &   0.01 \\   
15 & 2.38  &   1.70  &  1.59  &  1.76  &  1.78   &  1.77    &  0.36    &  0.10   &   0.01 \\   
16 & 2.19  &   1.77  &  1.79  &  1.93  &  1.85   &  1.95    &  0.17    &  0.03   &   0.04 \\   
17 & 2.19  &   1.77  &  1.79  &  1.93  &  1.85   &  1.95    &  0.17    &  0.03   &   0.04 \\   
18 & 2.19  &   1.77  &  1.79  &  1.93  &  1.85   &  1.95    &  0.17    &  0.03   &   0.04 \\   
19 & 2.19  &   1.77  &  1.79  &  1.93  &  1.85   &  1.95    &  0.17    &  0.03   &   0.04 \\   
20 & 1.84  &   1.82  &  1.67  &  1.76  &  1.67   &  1.78    &  0.02    &  0.03   &   0.03 \\   
21 & 1.84  &   1.82  &  1.67  &  1.76  &  1.67   &  1.78    &  0.02    &  0.03   &   0.03 \\
22 & 2.11  &   1.66  &  1.72  &  1.83  &  1.79   &  1.85    &  0.16    &  0.07   &   0.04 \\
23 & 2.11  &   1.66  &  1.72  &  1.83  &  1.79   &  1.85    &  0.16    &  0.07   &   0.04 \\
   \hline
   \end{tabular}
\end{table*}  
\section {Acknowledgments} Financial support from the Deutsche Forschungsgemeinschaft (DFG) within the Schwerpunktprogramm SPP1458 is gratefully acknowledged. L.J. thanks the Alexander-von-Humboldt foundation for financial support. We thank A.~Akbari, P. Hirschfeld, and P.~Thalmeier for discussions. We specially thank I.~Eremin for discussions as well as comments on the manuscript. 
%


\end{document}